\documentclass[epj]{svjour}
\usepackage{graphics}

\begin{document}

\title{Striped antiferromagnetic order and electronic
properties of stoichiometric LiFeAs from first-principles
calculations}

\author{Yong-Feng Li and Bang-Gui Liu\thanks{e-mail: \texttt{bgliu@mail.iphy.ac.cn}}}

\institute{ Institute of Physics, Chinese Academy of Sciences,
Beijing 100190, China \\ Beijing National Laboratory for Condensed
Matter Physics, Beijing 100190, China}

\abstract{We use state-of-the-arts first-principles method to
investigate the structural, electronic, and magnetic properties of
stoichiometric LiFeAs. We optimize fully all the structures,
including lattice constants and internal position parameters, for
different magnetic orders. We find the magnetic ground state by
comparing the total energies among all the possible magnetic
orders. Our calculated lattice constants and As internal position
are in good agreement with experiment. The experimental fact that
no magnetic phase transition has been observed at finite
temperature can be attributed to the tiny inter-layer spin
coupling. Our results show that stoichiometric LiFeAs has almost
the same striped antiferromagnetic spin order as other FeAs-based
parent compounds and tetragonal FeSe do, which may imply that all
Fe-based superconductors have the same mechanism of
superconductivity.}


\titlerunning{Striped antiferromagnetic order and
electronic properties of stoichiometric LiFeAs}
\authorrunning{Yong-Feng Li et al.}
 \maketitle

\section{Introduction}

Since the discovery of superconductivity in F-doped LaFeAsO
\cite{a1}, many Fe-based superconductors have been achieved by
doping appropriate atoms in or applying pressure on $R$OFeAs ($R$:
rare earth) \cite{a2,a3,a4}, $A$Fe$_2$As$_2$ ($A$: alkaline earth)
\cite{b1,b2,b3,1}, SrFeAsF \cite{Sr1,Sr2,Sr3,Sr4}, LiFeAs
\cite{5,8,7,1}, and even FeSe \cite{FeSe1,FeSe2,FeSe3,FeSe4}. The
highest transition temperature $\mathrm{T_c}$ has already reached
to 55-56 K \cite{a4}. Various studies have been performed to
understand their magnetic orders, electronic structures,
superconductivity, etc~\cite{a5,a6,a7,a8,a9,a10,a11,3}. Among the
five series of Fe-based superconductors, the LiFeAs series seems
to be the simplest FeAs-based superconductors. Many experiments
have been done to promote
$\mathrm{T_c}$\cite{4,5,6,7,8,9,10,12,13}. It should be easier to
detect the mechanism of the superconductivity in this series.
Although there have been some computational studies on
LiFeAs\cite{1,2,3}, many aspects, even the magnetic properties of
stoichiometric LiFeAs, are not elucidated. It is highly desirable
to solve such basic issues of LiFeAs for further investigations
and future applications of the FeAs-based materials.

Here, we investigate the structural, electronic, and magnetic
properties of stoichiometric LiFeAs by using state-of-the-arts
first-principles calculations. We compare all the possible
magnetic orders, and find the magnetic ground state in terms of
total energy results and force optimization. Our results show the
magnetic ground state features a striped antiferromagnetic (SAF)
order in the Fe layer and a weak antiferromagnetic (AFM) order in
the $z$ direction perpendicular to the Fe layer. Our calculated
internal positions of Li and As are in good agreement with
experiment. Through analyzing electronic and the magnetic results,
we propose a simple spin model to understand the experimental
magnetic properties. Thus, we show that stoichiometric LiFeAs has
almost the same SAF spin order as other FeAs-based parent
compounds and tetragonal FeSe do. More detailed results will be
presented in the following.

In next section we shall describe our computational details. In
the third section we shall optimize fully all the structures with
different magnetic orders and thus determine the magnetic ground
state. In the fourth section we shall present the electronic
structures of the ground state phase. Finally, we shall make some
discussions and give our conclusion in the fifth section.

\section{Computational details}

We investigate all possible magnetic orders for stoichiometric
LiFeAs by using full-potential linearized augmented plane wave
method within the density functional theory \cite{dft}, as
implemented in package WIEN2k\cite{wien2k,wien2ka}. The
generalized gradient approximation is used for the exchange and
correlation potentials\cite{pbe96}. We treat Li-2s, Fe-3d4s, and
As-4s4p as valence states, Fe-3p and As-3d as semi-core states,
and the lower states as the core states. For the core states, the
full relativistic effect is calculated with radial Dirac
equations. For valence and semi-core states, the relativistic
effect is calculated under the scalar approximation, with
spin-orbit interaction being neglected\cite{relsa}. We take
$\mathrm{R_{mt}\times K_{max}}$=7.5 and make the angular expansion
up to $l=10$ in the muffin-tin spheres. We use 1000 k points in
the first Brillouin zone for different magnetic structures. The
self-consistent calculations are controlled by the charge density
in real space. The integration of the absolute charge-density
difference between two successive self-consistent loops, as the
convergence standard, is less than 0.0001$|e|$ per unit cell,
where $e$ is the electron charge. For all the magnetic structures,
we optimize the volume, geometry, and internal position parameters
in terms of total energy and the force by the standard 1.5
mRy/a.u.

\begin{figure}
\begin{center}
\scalebox{0.3}{\includegraphics{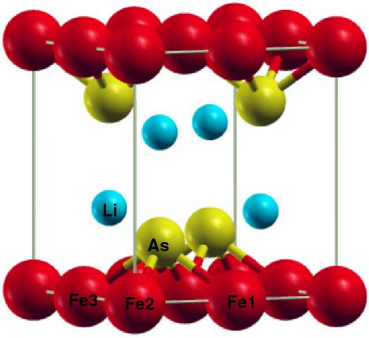}} \caption{(color
online). Schematic crystal structure of stoichiometric LiFeAs
(P4/nmm). The biggest ball is Fe atom, the medium As, and the
smallest Li. }
\end{center}
\end{figure}

\begin{table*}
\begin{center}
\caption{The relative total energy per unit cell ($E$, with that
of the lowest SAF-AFM set to zero), the magnetic moment of Fe atom
($M$), the equilibrium volume  per unit cell ($V$), the internal
position parameters of Li and As ($z_{_\mathrm{Li}}$ and
$z_{_\mathrm{As}}$), the distance between Fe and As atoms $d_{\rm
Fe-As}$, two angles formed by Fe1-As-Fe2 ($\alpha$) and Fe1-As-Fe3
($\beta$) of LiFeAs in the six possible magnetic orders.}
\begin{tabular*}{0.80\hsize}{cccccccccc} \hline\hline
Name & $E$ (meV)& $M$ ($\mu_{B}$) & $V$ ($\rm{\AA^3}$) & $z_{\rm Li}$ &  &$z_{\rm As}$&$d_{\rm Fe-As}$ (\AA)&$\alpha$&$\beta$\\
FM & 140. & 0.43 & 88.47 & 0.3304 & & 0.2238&2.345&68.8&106.1\\
NM & 151. &--&87.85&0.3313 & & 0.2237&2.339&68.9&106.1\\
CAF-AFM & 151. & 1.02 & 87.83 & 0.3330 & & 0.2284&2.356&68.4&105.4\\
CAF-FM & 148. & 1.18 & 88.45 & 0.3330 & & 0.2286&2.362&68.3&105.0\\
SAF-FM & 2. &1.50&88.31&0.3353& & 0.2324&2.376&67.8&104.0\\
SAF-AFM & 0 &1.58&88.49&0.3322& & 0.2334&2.382&67.6&103.8\\
\hline\hline
\end{tabular*}\end{center}
\end{table*}

\section{Fully optimized structures with different magnetic orders}

Stoichiometric LiFeAs crystallizes into a tetragonal structure with
space group P4/nmm at room temperature. As is shown In Fig. 1, its
unit cell consists of one Fe-As layer separated by two Li layers.
The experimental lattice constants are $a=3.791\rm{\AA}$ and
$c=6.364\rm{\AA}$)\cite{8}. To determine the ground state of LiFeAs,
we arrange the Fe moments in the Fe layer so as to form nonmagnetic
(NM), ferromagnetic (FM), checkerboard antiferromagnetic (CAF), and
SAF orders. In the $z$ direction, the successive Fe layers can
couple ferromagnetically or antiferromagnetically. Thus, we have
four different AFM orders, namely, CAF-FM, CAF-AFM, SAF-FM and
SAF-AFM. Structural optimization is done for these six magnetic
orders. Our volume optimization results are presented in Fig. 2. The
equilibrium volume is determined by the minimum of the total energy
against volume. The optimized results summarized Table I. The Fe
layer in SAF order, either FM or AFM in $z$ direction, is lower than
that in other magnetic orders. We conclude that the magnetic ground
state of LiFeAs is the magnetic configuration SAF-AFM that the spins
in the Fe layers are in the SAF order and the inter-layer magnetic
coupling is a weak AFM interaction. This is the same as the magnetic
ground states of other FeAs-based parent compounds and tetragonal
FeSe\cite{1,Sr4,FeSe3,a7,a8}.

The equilibrium volume (88.49\AA$^3$) of the ground state is a
little smaller than the experiment value (91.46 $\rm {\AA^3}$).
The magnetic moment per Fe atom is 1.58 $\mu_B$. The moment for
the CAF order is smaller. For comparison, we do similar
calculations using LSDA. The LSDA magnetic moment per Fe atom,
0.62$\mu_B$ and 0.63$\mu_B$ for the SAF-FM and SAF-AFM orders, are
much smaller than the GGA results (1.50$\mu_B$ and 1.58$\mu_B$),
respectively. The Fe-As layer, especially the Fe-As distance and
the Fe-As-Fe angle, is important for the ground state of LiFeAs.
This can be easily seen in Table 1 that as the magnetic moment $M$
of Fe atom increases, the values of the distance $d_{\rm Fe-As}$
between Fe and As atoms increases and  the Fe-As-Fe angles
($\alpha$ and $\beta$) decreased. The trend of $M$ with different
$d_{\rm Fe-As}$ is consistent with previous result calculated with
experimental lattice constants \cite{3}. The Fe-As layer as a
whole is quite robust and Li atoms are dispersed independently
between the Fe-As layers, which explains why we can get exact
positions of As atoms (0.2334 compared to the experiment value
0.2365 ), but a little deviated results for Li atoms (0.3322 to
0.3459). It looks like that the Li atoms play less important roles
in the materials.

\begin{figure}
\begin{center}
\resizebox{0.9\hsize}{!}{\includegraphics*{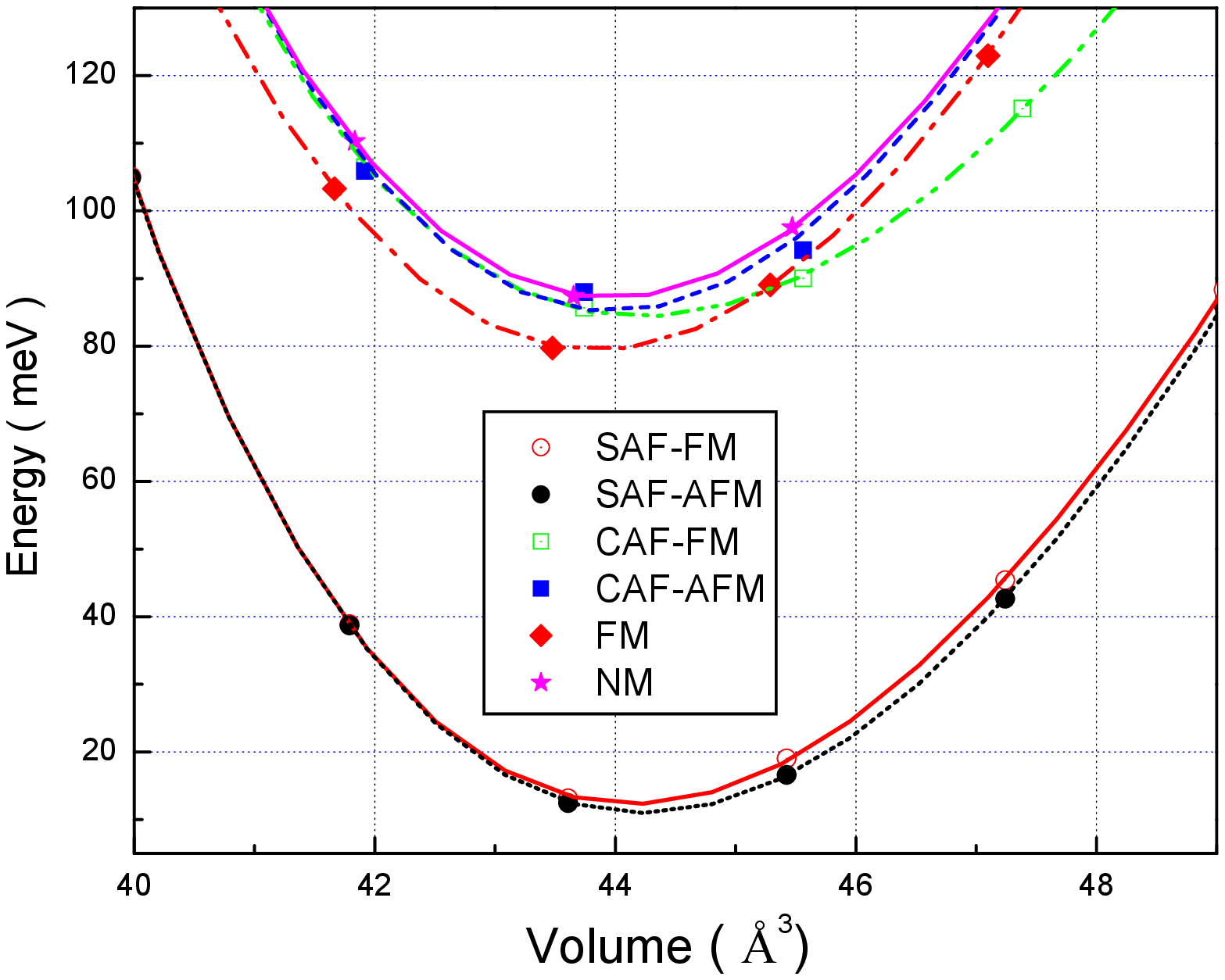}}
\caption{(color online). The optimization curves of relative total
energy (meV) as functions of the volume ($\rm{\AA^3}$), per
formula unit, for the six possible magnetic orders of
stoichiometric LiFeAs.} \end{center}
\end{figure}

\section{Spin dependent Electronic structures}

\begin{figure}
\begin{center}
\resizebox{0.8\hsize}{!}{\includegraphics*{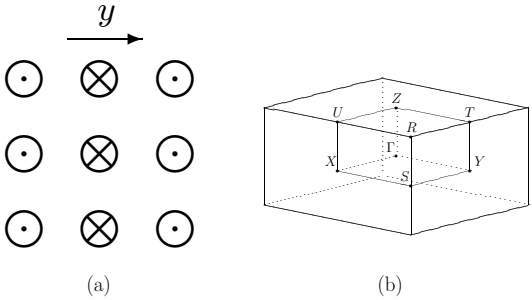}} \caption{ (a)
The magnetic structure of the Fe layer of the LiFeAs ground state,
(b) The first brillouin zone of the LiFeAs ground-state phase with
the high-symmetry points labelled.} \end{center}
\end{figure}

In the following, we study the electronic properties of
stoichiometric LiFeAs in SAF order. The magnetic structure in the
Fe layer is shown in Fig. 3(a). The arrow implies that the Fe
spins align ferromagnetically in the $x$ direction and
antiferromagnetically in $y$ direction. Also we show the first
Brillouin zone with key representative points and lines emphasized
in Fig. 3(b). The total spin-dependent density of states (DOS) and
the partial DOSs projected in the muffin-tin spheres of Fe1, Fe2,
Li, and As atoms and in the interstitial region are shown in Fig.
4. There is no energy gap near the Fermi level and then it shows a
metal feature. For the SAF order, the spins of Fe1 and Fe2 are
antiparallel, it can be easily recognized in DOS. The DOS can be
divided into two energy ranges: from -6.0 eV to -3.0 eV  and from
-3.0 eV to 0 eV (Fermi level). The former consists of Fe-3d and
As-4p states forming the Fe-As bond. In the latter range, the
Fe-3d states play a key role. It is easily noted that the Li
states have very small weight in the energy window. The spin
exchange splitting should be mediated by the d-d direct exchange
between the nearest Fe atoms and the p-d hybridization between Fe
and As.

\begin{figure}[!htb]
\begin{center}
\resizebox{0.95\hsize}{!}{\includegraphics*{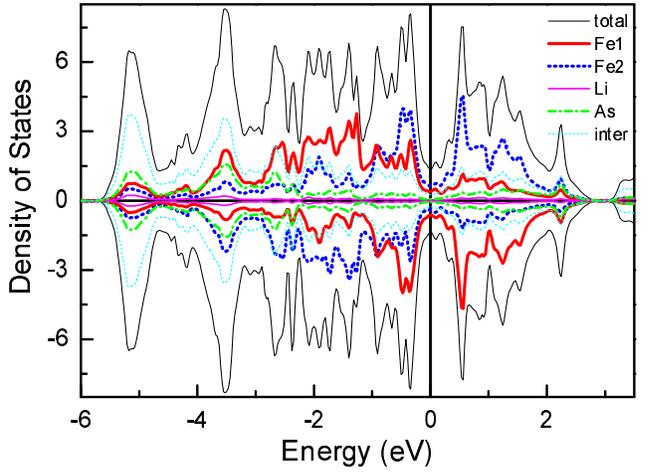}}
\caption{(color online). Total DOS (states/eV per formula unit)
and partial DOS projected in the atomic spheres of Fe1, Fe2, Li,
and As and in the interstitial region of the LiFeAs ground-state
phase. The upper part is for majority spin and the other for
minority spin.}
\end{center}
\end{figure}

In Fig. 5 we present the spin-dependent band structure of LiFeAs
with Fe1-3d character in SAF order. The plots look like lines, but
consist of hollow circles. The circle diameter is proportional to
the weight of Fe1-3d states at that $k$ point. Compared with DOS,
we can find that 3d states of Fe play a key role for the magnetic
structure. It is clear that the bands become narrow along the $z$
direction. It means that the electron structure shows quite strong
two-dimensional character. Also, it can be found that there are
hole-like (along $\Gamma$-Z) and electron-like (along Z-U) band
sections.

The crystal structures with NM, FM, and CAF orders have the same
tetragonal symmetry, while that with SAF is orthorhombic.
Different symmetries lead to two different kinds of crystal field
splitting for the Fe-3d state. For the tetragonal symmetry, Fe-d
state is split into d$_{z^2}$, d$_{xy}$, d$_{x^2-y^2}$, and
d$_{xz,yz}$, and for orthorhombic symmetry it is split into five
singlets. From Fig. 5 we can see that the bands distributions are
symmetric along $\Gamma$-X and $\Gamma$-Y at -3 eV below the Fermi
level, but asymmetric near the Fermi level. These are different
from the previous NM band structures\cite{1,3,2}, in which the
band distributions are symmetric along $\Gamma$-X and $\Gamma$-Y
for almost all energy range below the Fermi level.

\begin{figure}[!htb]
\begin{center}
\resizebox{0.95\hsize}{!}{\includegraphics*{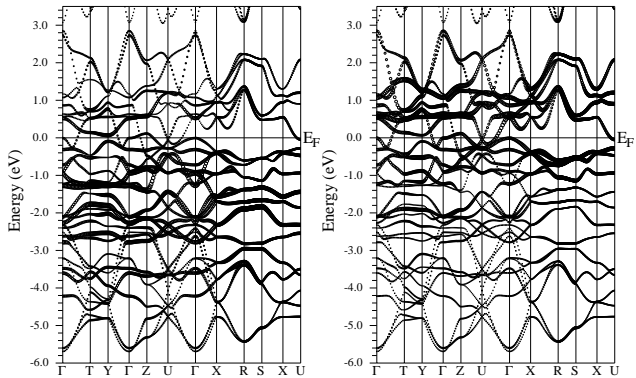}}\caption{The
spin-dependent energy bands of the LiFeAs ground-state phase. The
left part is for majority-spin and the right for minority-spin.
The band consists of dots. The bigger the dot is, the more Fe1-3d
character the band at that point has. }
\end{center}
\end{figure}

\section{Discussions and conclusion}

The total energy results, as summarized in Fig. 2 and Table I,
show that the magnetic interaction in the $z$ direction is weak,
and The SAF structure is lower by 140-151 meV per formula unit
than the other magnetic configurations. This is consistent with a
simple spin model for the Fe spins, $ H = \sum_{\langle
ij\rangle}J_{ij}\vec{S}_i\cdot\vec{S}_j $, where $\vec{S}_i$ is
the spin operator at site $i$ and $J_{ij}$ coupling constants
between sites $i$ and $j$. The summation is limited to the nearest
site pairs. $J_{ij}$ is $J$ in the $y$ direction, $-J$ in the $x$
direction, and $\delta J$ in the $z$ direction. For such spin
model, a non-zero $\delta$ is necessary to a finite phase
transition temperature\cite{lbg}. The actual $\delta$ must be tiny
because no magnetic phase transition is observed in stoichiometric
LiFeAs at any finite temperature.

In summary, we have investigated the structural, electronic, and
magnetic properties of stoichiometric LiFeAs by using the
density-functional-theory method. We determine the magnetic
ground-state by comparing the total energies among all the
possible magnetic orders. Our DFT results show the magnetic ground
state has the SAF order in the Fe layer and a weak AFM order in
the $z$ direction. Our calculated internal positions of Li and As
are in good agreement with experiment. Through analyzing
electronic and the magnetic results, we propose a simple spin
model to understand the experimental magnetic properties. The
experimental fact that no magnetic phase transition is observed at
finite temperature can be attributed to the tiny inter-layer spin
coupling. Thus, we show that stoichiometric LiFeAs has almost the
same SAF spin order as other FeAs-based parent compounds and
tetragonal FeSe do\cite{1,Sr4,FeSe3,a7,a8}, which may imply that
all Fe-based superconductors have the same mechanism of
superconductivity.

\begin{acknowledgement}
Acknowledgement. This work is supported  by Nature Science
Foundation of China (Grant Nos. 10874232 and 10774180), by the
Chinese Academy of Sciences (Grant No. KJCX2.YW.W09-5), and by
Chinese Department of Science and Technology (Grant No.
2005CB623602).
\end{acknowledgement}

\end{document}